\documentclass[a4paper,11pt]{article}
\usepackage{dcolumn}
\usepackage{bm}
\usepackage[usenames,dvipsnames]{color}
\usepackage{graphicx,amssymb,amsmath,amsfonts,epsfig}
\usepackage{textcomp}
\usepackage{float}

\title{Thermodynamic Volume Product in Spherically Symmetric and Axisymmetric Spacetime}
\author{Parthapratim Pradhan\footnote{pppradhan77@gmail.com}  
\\ {\it  Department of Physics }
\\ {\it Hiralal Mazumdar Memorial College For Women }
\\ {\it Dakshineswar, Kolkata-700035, India }
}
\date{}

\begin{document}

\maketitle

\abstract{In this Letter, we have examined the thermodynamic volume products for spherically symmetric and 
axisymmetric spacetimes in the framework of \emph{extended phase space}. Such volume products usually formulated 
in terms of the outer horizon~(${\cal H}^{+}$) and the inner horizon~(${\cal H}^{-}$) of black hole
~ {(BH)} spacetime. 
Besides volume product, the other thermodynamic formulations like \emph{volume sum, volume minus and volume division} 
are considered for a wide variety of spherically symmetric spacetime and axisymmetric spacetimes. 
Like area~(or entropy) product of multihorizons, the mass-independent~(universal) feature of volume products 
{are} sometimes also \emph{fail}. In particular for a spherically symmetric AdS spacetimes
the simple thermodynamic 
volume product of ${\cal H}^{\pm}$ is not mass-independent.  In this case,  more complicated combinations of 
outer and inner horizon volume products are indeed mass-independent. For a particular class of spherically 
symmetric cases i.e.  Reissner Nordstr\"{o}m { BH} of Einstein gravity and Kehagias-Sfetsos
{BH} of 
Ho\v{r}ava Lifshitz gravity, the thermodynamic volume products of ${\cal H}^{\pm}$ is indeed \emph{universal}. 
For axisymmetric class of {BH } spacetime in Einstein gravity all the  combinations 
are \emph{mass-dependent}. 
There has been no chance to formulate any combinations of volume product relation is to be mass-independent. 
Interestingly, \emph{only the rotating BTZ black hole} in  3D provides the volume product formula is 
mass-independent i.e. \emph{universal} and hence it is  quantized.}


\section{Introduction}
It has been examined by a number of researchers that the area (or entropy) product of various spherically symmetric and 
axisymmetric {BHs } are mass-independent (universal)~\cite{ah09,cgp11,castro12,castro13,sd12,mv13,jh14,pp14,pp15}.  
For instance, Ansorg and Hennig~\cite{ah09} demonstrated  that for a stationary and axisymmetric class of Einstein-Maxwell
gravity the area product formula satisfied the universal relation as 
\begin{eqnarray}
{\cal A}_{h} {\cal A}_{c} &=& (8\pi J)^2+(4\pi Q^2)^2   ~.\label{v1}
\end{eqnarray}
where ${\cal A}_{h}$ and ${\cal A}_{c}$ are area of outer horizon~(OH) or event horizon~(EH) and inner horizon~(IH) or 
Cauchy horizon~(CH). The parameters, $J$ and $Q$ are denoted as the angular momentum and charge of the black hole~(BH) 
respectively. 

On the other hand, Cveti\v{c} et al.~\cite{cgp11} extended this work for a higher dimensions spacetime and showed 
that for multihorizon BHs the area product formula should be quantized by satisfing the following relation
\begin{eqnarray}
{\cal A}_{h} {\cal A}_{c}  &=&   (8 \pi \ell _{pl}^2)^2 N , \,\, N\in {\mathbb{N}} ~.\label{v2}
\end{eqnarray}
where $\ell _{pl}$ is the Planck length. This relation indicates that the product relation is indeed universal in nature.
This is a very fascinatic topic of research since 2009.

Aspects of BH thermodynamic properties have started by the seminal work of Hawking and Page~\cite{haw83}, they first 
proposed that certain type of phase transition occurs between small and large BHs in case of 
Schwarzschild-AdS BH. This phase transition is now called the famous Hawking-Page phase transition. For a charged 
AdS BH, the study of thermodynamic properties initiated by Chamblin et al.~\cite{chamblin99,chamblin99a}, where the 
authors demonstrated that the critical behaviour of Van-der Waal like liquid-gas phase transitions. This has been 
brought into a new form by Kubiz\v{n}\'{a}k and Mann~\cite{david12} by examining the thermodynamic properties 
i.e. $P-V$ criticality of Reissner-Nordstr\"{o}m AdS BH in the extended phase space. They determined the BH equation 
of state and computed the critical exponent by using the mean field theory and also computed the other thermodynamic 
features.

Motivated by the above mentioned work and our previous investigation~\cite{ppmpd} {in which} we have 
considered the \emph{extended phase space} framework for a wide variety of spherically symmetric AdS spacetime. 
In the present work, we would like to extend our study for {various classes of spherically symmetric BHs
and axisymmetric BHs}. 
In {the extended phase formalism}, the cosmological constant is treated as  thermodynamic pressure 
$P$ and its conjugate variable as  thermodynamic volume ${\cal V}$~\cite{kastor09,dolan10,dolan11,david12}. 
They {are} defined as 
\begin{eqnarray} 
P &=& -\frac{\Lambda}{8\pi}=\frac{3}{8\pi \ell^2}  ~.\label{pr}
\end{eqnarray}
and  
\begin{eqnarray}
{\cal V} &=& \left(\frac{\partial M}{\partial P}\right)_{S,Q,J}  ~.\label{vm}
\end{eqnarray} 
{The extended phase space is more meaningful than conventional phase space due to the following reasons.
The conventional phase space allows the physical parameters like temperature, entropy, charge and potential etc. .
Whereas the extended phase space allows the parameters like pressure, volume and enthalpy~(rather than internal energy). 
In addition to that the mass parameter should be considered there as enthalpy of the system, which is useful to study 
the critical behaviour of the thermodynamic system. The BH equation of state could be used to study for comparisons 
with the classical thermodynamic  equation of state~(Van der-Waal equation). Once the BH thermodynamic equation of 
state is in hand then one may compute different thermodynamic quantities like isothermal compressibility, specific 
heat at constant pressure etc. 
}

This thermodynamic volume~\footnote{There are different types of definitions regarding the volume of a BH in the literature. 
The idea regarding the BH volume was first introduced by Parikh~\cite{parikh}. For other types of definition like dynamical 
volume and vector volume (see Ref. ~\cite{roveli,lake,matzner}).  Here we are particulaly interested regarding the 
thermodynamic volume ~\cite{cvetic11}.}  of a spherically symmetric BH  and for OH should read 
\begin{eqnarray}
{\cal V}_{h} &=& \frac{4}{3}\pi r_{h}^3=\frac{{\cal A}_{h} r_{h}}{3}  ~.\label{vm1}
\end{eqnarray} 
where $r_{h}$ is OH radius. Similarly, this volume for IH should be 
\begin{eqnarray}
{\cal V}_{c} &=& \frac{4}{3}\pi r_{c}^3=\frac{{\cal A}_{c} r_{c}}{3}  ~.\label{vmc}
\end{eqnarray} 

It should be noted that the thermodynamic volume of CH can be obtain by using the symmetric 
properties~\cite{ppmpd} of OH radius $r_{h}$  and IH radius $r_{c}$ i.e.
\begin{eqnarray}
{\cal V}_{c} = V_{h}|_{r_{h}\leftrightarrow r_{c}} ~.\label{ioh}
\end{eqnarray}

Another important point in the extended phase space is that the ADM mass should be treated as 
the total  enthalpy of the  thermodynamic system i.e. $M=H=U+P{\cal V}$. Where $U$ is thermal energy of the 
system~\cite{kastor09}. Therefore the first law of BH thermodynamics in this phase space for any 
spherically symmetric spacetime and for OH should be
\begin{eqnarray}
dH &=&  T_{h} dS_{h} + {\cal V}_{h} dP +\Phi_{h} dQ ~. \label{vm3}
\end{eqnarray}
where the quantities $T_{h}$, $S_{h}$ and $\Phi_{h}$ are denoted as the BH temperature, entropy 
and electric potential of OH. The parameter $Q$ is denoted as the charge of a BH.

Analogously, the first law of BH mechanics  for IH should be
\begin{eqnarray}
dH &=& - T_{c} dS_{c} + {\cal V}_{c} dP +\Phi_{c} dQ ~. \label{vmc3}
\end{eqnarray}
where the quantities $T_{c}$, $S_{c}$ and $\Phi_{c}$ are denoted as the  corresponding 
BH temperature, entropy and electric potential which could be defined  on the IH. 

When we add the rotation parameter the first law of BH thermodynamics in the extended phase space 
(for axisymmetric spacetime and for OH) becomes 
\begin{eqnarray}
dH &=&  T_{h} dS_{h} + {\cal V}_{h} dP +\Phi_{h} dQ +\Omega_{h}dJ ~. \label{vm5}
\end{eqnarray}
where $\Omega_{h}$ and $J$ are the angular velocity defined on the OH and the angular momentum of BH. 
For IH, the first law becomes
\begin{eqnarray}
dH &=& - T_{c} dS_{c} + {\cal V}_{c} dP +\Phi_{c} dQ +\Omega_{c} dJ ~. \label{vmc5}
\end{eqnarray}
where $\Omega_{c}$ is the angular velocity defined on the IH. Using symmetric features of $r_{h}$ and $r_{c}$, one 
can determine the following thermodynamic relations for IH
\begin{eqnarray}
{\cal A}_{c} &=& {\cal A}_{h}|_{r_{h}\leftrightarrow r_{c}}, {\cal S}_{c}={\cal S}_{h}|_{r_{h}\leftrightarrow r_{c}}, 
\Omega_{c}=\Omega_{h}|_{r_{h}\leftrightarrow r_{c}}, \Phi_{c}=\Phi_{h}|_{r_{h}\leftrightarrow r_{c}} \nonumber\\
T_{c} &=& -T_{h}|_{r_{h}\leftrightarrow r_{c}}, \, {\cal V}_{c} = {\cal V}_{h}|_{r_{h}\leftrightarrow r_{c}}
~.\label{re1}
\end{eqnarray}
However in this work, we wish to extend our study by computing the volume product, volume sum, volume minus and 
volume division in the \emph{extended phase space} for various spherically symmetric BHs and axisymmetric BHs~
(including {the various} AdS spacetimes). By evaluating these quantities we prove that for 
a spherically symmetric AdS spacetime
the simple volume product is \emph{not} mass-independent.  In this case, somewhat complicated combination of volume
functional relations of OH and IH are indeed mass-independent. For instance, we have derived  the 
mass-independence volume functional relation for RN-AdS BH as
\begin{eqnarray}
f({\cal V}_{h}, {\cal V}_{c}) &=& \ell^2 ~, \label{in}
\end{eqnarray}
where
\begin{eqnarray}
f({\cal V}_{h}, {\cal V}_{c}) &=&
\left(\frac{3}{32 \pi}\right)^{\frac{1}{3}} \frac{\left(\frac{8\pi\ell^2 Q^2}{3}\right)}
{({\cal V}_{h} {\cal V}_{c})^{\frac{1}{3}}}-\left(\frac{3}{4\pi}\right)^{\frac{2}{3}}
\left[{{\cal V}_{h}}^{\frac{2}{3}}+{{\cal V}_{c}}^{\frac{2}{3}}+({\cal V}_{h}{\cal V}_{c})^{\frac{1}{3}}
\right] ~. \label{inn}
\end{eqnarray}
For  simple Reissner Nordstr\"{o}m BH~(which is a spherically symmetric solution of Einstein equation) of 
Einstein gravity and Kehagias-Sfetsos BH of Ho\v{r}ava Lifshitz gravity, the thermodynamic volume products 
of ${\cal H}^{\pm}$ is mass-independent. Therefore they behaves as a universal character by its own features. 
Moreover, we have derived the thermodynamic volume functional relation for  Ho\v{r}ava Lifshitz-AdS BH and 
phantom AdS BH. The phantom fields are exotic because it was produced via negative energy density. Furthermore 
we have derived volume functional relation for regular BH. Regular BH is a kind of BH which is free from a curvature
singularity.

Whereas for axisymmetric class of BHs including AdS spacetimes there has been no chance to formulate any 
possible combinations of thermodynamic volume product is to be mass-independent. It should be noted that 
for a KN-AdS BH, there may be a possibility to formulate the area~(or entropy) product relations are 
mass-independent. The reason is that for a simple Kerr BH, the area~(or entropy) product is universal i.e. 
mass-independent while the \emph{volume product} is not! Because the thermodynamic volume is proportional 
to the spin parameter. That's why \emph{there has been no chance to produce any combinations of volume product}
of ${\cal H}^{\pm}$ is mass-independent. Therefore the axisymmetric BHs showing that \emph{no} universal behaviour 
for volume products. Interestingly, only rotating BTZ BH shows that the mass-independent feature. 
Thus \emph{only} axisymmetric BHs in 3D provided that the universal character of thermodynamic volume product.

In our previous investigation~\cite{pp14,pp15}, we computed the BH area (or entropy) 
{products, BH temperature products,
Komar energy products and specific heat products for various classes of BHs}.  Besides the area (or
entropy) product, it should be important to study whether the thermodynamic \emph{volume product, volume sum, volume minus 
and volume division} for all the horizons are universal or not and whether 
they should be quantized or not. This is the main motivation behind this work. 

The structure of the paper is as follows. In Sec.~2, we shall compute the various thermodynamic volume products for spherically 
symmetric BHs and conclude that the product is mass-independent.  In Sec.~3, we compute various thermodynamic volume products 
for axisymmetric spacetime and conclude that the product is mass-dependent. Interestingly, for the spinning BTZ BH the said 
volume product is \emph{mass-independent}. 

\section{Spherically Symmetric BH}
In this section we {would} consider various spherically symmetric BHs.

\subsection{Reissner Nordstr\"{o}m BH}
We begin with charged BH with zero cosmological constant which is a solution of Einstein equation. The metric 
form is given by 
\begin{eqnarray}
ds^2 &= & -{\cal Z}(r) dt^2 + \frac{dr^2}{{\cal Z}(r)} +r^2 d\Omega_{2}^2 .~\label{met}
\end{eqnarray}
where,
\begin{eqnarray}
{\cal Z}(r) &=&  1-\frac{2M}{r}+\frac{Q^2}{r^2},~\label{h2}
\end{eqnarray}
and $d\Omega_{2}^2$ is metric on the unit sphere in two dimensions.

The OH~\footnote{There are several definitions of horizons for a  static spherically  symmetric spacetime. 
We have used Killing horizons for computaions of thermodynamic volume.} radius and IH radius reads 
\begin{eqnarray} 
r_{h}=M + \sqrt{M^2-Q^2} \\
r_{c}= M - \sqrt{M^2-Q^2}  ~.\label{rn1}
\end{eqnarray}
where $M$ and $Q$ denotes the mass and charge of BH respectively.
When $M^2>Q^2$, it descibes a BH otherwise it has a naked singularity. The thermodynamic volume for OH  and IH  
should read 
\begin{eqnarray}
{\cal V}_{h} &=& \frac{4}{3}\pi r_{h}^3  \\
{\cal V}_{c} &=& \frac{4}{3}\pi r_{c}^3 ~.\label{rn2}
\end{eqnarray} 
The thermodynamic volume~\footnote{In the limit $Q=0$, one obtains the thermodynamic volume for Schwarzschild BH. 
Since in this case the BH has only OH  located at $r_{h}=2M$. Therefore the volume should be 
${\cal V}_{h}= \frac{32}{3}\pi M^3$. Thus for an isolated Schwarzschild BH the thermodynamic volume should be 
mass dependent therefore it is not universal and not quantized in nature by its own character.} 
product for OH and IH  should be 
\begin{eqnarray}
{\cal V}_{h} {\cal V}_{c} &=& \frac{16}{9}\pi^2 Q^6   ~.\label{rn3}
\end{eqnarray} 
It is indeed mass-independent thus it is universal in character and it is also quantized. 

The volume sum for OH  and IH  is calculated to be 
\begin{eqnarray}
{\cal V}_{h}+ {\cal V}_{c} &=& \frac{32}{3}\pi M^3 \left(1-\frac{3}{4} \frac{Q^2}{M^2} \right)   ~.\label{rn4}
\end{eqnarray} 
Similarly, one can compute the volume minus for OH  and IH  as 
\begin{eqnarray}
{\cal V}_{h}- {\cal V}_{c} &=& \frac{32}{3}\pi M^2 \sqrt{M^2-Q^2} \left(1- \frac{Q^2}{4M^2} \right)   ~.\label{rn5}
\end{eqnarray} 
and the volume division should be 
\begin{eqnarray}
\frac{{\cal V}_{h}}{ {\cal V}_{c}} &=& \left(\frac{M + \sqrt{M^2-Q^2}}{M -\sqrt{M^2-Q^2}} \right)^3   ~.\label{rn6}
\end{eqnarray} 
It follows from the calculation that all these quantities are mass dependent so they are not universal 
in nature by its own right. From Eq.~(\ref{rn5}) and Eq.~(\ref{rn6}), we can easily see that in the extremal 
limit $M^2=Q^2$, one obtains ${\cal V}_{h}= {\cal V}_{c}$. This is a new condition of extreme limit in spherically 
symmetric cases.

\subsection{Ho\v{r}ava Lifshitz BH}
{In this section, we would briefly review the UV complete theory of gravity which is a
non-relativistic renormalizable theory of gravity known as Ho\v{r}ava Lifshitz~\cite{ph9a,ph9b,ph9c} gravity. It 
reduces to Einstein's gravity at large scales for the value of dynamical coupling constant~$\lambda=1$. Using 
ADM formalism, one could write the metric as
\begin{eqnarray}
ds^2 &= & -N^2 dt^2 +g_{ij}(dx^{i}-N^{i}dt)(dx^{j}-N^{j}dt)~.\label{hg1}
\end{eqnarray}
In addition for a spacelike hypersurface with a fixed time the extrinsic curvature $K_{ij}$ is given by
\begin{eqnarray}
K_{ij} &= &\frac{1}{2N}(\dot{g_{ij}}-\nabla_{i}N_{j}-\nabla_{j}N_{i})~.\label{hg2}
\end{eqnarray}
where a dot represents a derivative with respect to $t$.
The generalized action for Ho\v{r}ava Lifshitz could be written as
$$
S = \int dt d^3x\sqrt{g}N \big[ \frac{2}{\kappa^2}\big( K_{ij}K^{ij}-\lambda K^{2} \big)+
\frac{\kappa^2 \mu^2(\Lambda_{w} R-3\Lambda^2_{w})}{8(1-3\lambda)}+
\frac{\kappa^2 \mu^2(1-4\lambda)}{32(1-3\lambda)}R^2
$$
\begin{eqnarray}
-\frac{\kappa^2}{2w^4}\big(C_{ij}-\frac{\mu w^2}{2}R_{ij} \big)
\big(C^{ij}-\frac{\mu w^2}{2}R^{ij} \big)+\mu^4 R \big]~.\label{hg3}
\end{eqnarray}
Here $\kappa^2, \lambda, \mu, w$ and $\Lambda$ are the constant parameters and the cotton
tensor, $C_{ij}$ is defined to be
\begin{eqnarray}
C^{ij} &=& \epsilon^{ikl} \nabla_{k}\big(R^{j}_{l}-\frac{1}{4}\epsilon^{ikj}\partial_{k}R \big)
~.\label{hg4}
\end{eqnarray}
As compared with Einstein's general relativity, one could obtain the speed of light,
Newtonian constant and the cosmological constant as
\begin{eqnarray}
c &=& \frac{\kappa^2 \mu}{4}\sqrt{\frac{\Lambda_{w}}{1-3\lambda}} \\
G &=& \frac{\kappa^2}{32 \pi c} \\
\Lambda &=& \frac{3}{2}\Lambda_{w} ~.\label{hg4.1}
\end{eqnarray}
respectively. It should be mentioned here that when $\lambda=1$, the first three terms in Eq.\ref{hg3}
reduces to that one obtains as in   Einstein's gravity. It must also be noted that $\lambda$
is a dynamic coupling constant and for $\lambda >\frac{1}{3}$, the cosmological constant should be a
negative one. However, it could be made a positive one if one could give a following transformation like
$\mu\rightarrow i\mu$ and $w^2\rightarrow -iw^2$.
Here we restrict ourselves that the BH solution in the limit of $\Lambda_{w} \rightarrow 0$. That's why, 
we have to set $N^{i}=0$ and to get the spherically symmetric solution we have to choose the metric 
ansatz as
\begin{eqnarray}
ds^2 &= & -N^{2}(r) dt^2 + \frac{dr^2}{g(r)} +r^2(d\theta^2+ \sin^2\theta d\phi^2)~.\label{hg5}
\end{eqnarray}
In order to get the spherically symmetric solution, sustituting  the metric ansatz~\ref{hg5} into 
the action and one obtains reduced Lagrangian as
\begin{eqnarray}
{\cal L} &=& \frac{\kappa^2 \mu^2 N}{8(1-3\lambda) \sqrt{g}}\big[(2\lambda -1)\frac{(g-1)^2}{r^2}
-2\lambda \frac{g-1}{r}g'+ \frac{g-1}{2}g'^2 -2\omega(1-g-rg')  \big] ~.\label{as1}
\end{eqnarray}
where $\omega=\frac{8\mu^2(3\lambda-1)}{\kappa^2}$. Here we are interested
to investigate the situation $\lambda=1$ i.e. $\omega=\frac{16 \mu^2}{\kappa^2}$. 
Then one finds the solution of the metric~\cite{ks09} as
\begin{eqnarray}
N^2(r) &=& g = 1-\sqrt{4 M \omega r+\omega^2 r^4} +\omega r^2,\label{hl1}
\end{eqnarray}
where $M$ is an integration constant related to the mass parameter.
Thus the static, spherically symmetric solution  is given by
\begin{eqnarray}
ds^2 &= & -g(r) dt^2 + \frac{dr^2}{g(r)} +r^2(d\theta^2+ \sin^2\theta d\phi^2).~\label{hl}
\end{eqnarray}
For $r\gg (\frac{M}{\omega})^{\frac{1}{3}}$, one gets the usual behavior of a 
Schwarzschild BH. The BH horizons correspond to $g(r)=0$
}
The OH radius and IH radius should read
\begin{eqnarray} 
r_{h}=M + \sqrt{M^2 -\frac{1}{2\omega}} \\
r_{c}=M - \sqrt{M^2 -\frac{1}{2\omega}}
~.\label{ks}
\end{eqnarray}
where $M$ and $\omega$ denotes the mass and coupling constant of BH respectively.
When $M^2 >\frac{1}{2\omega}$, it descibes a BH and when $M^2 <\frac{1}{2\omega}$, it descibes a naked singularity.

The thermodynamic volume product for KS BH should be 
\begin{eqnarray}
{\cal V}_{h} {\cal V}_{c} &=& \frac{2\pi^2}{9\omega^3}   ~.\label{ks1}
\end{eqnarray} 
It indicates that it is mass-independent therefore it is universal in nature and it also be quantized. We do not 
calculate other possible combination because it is clear that these combinations are surely mass dependent as we 
have seen in case of  RN BH. It should be mentioned that 
the Smarr formula satisfied in case of Einstein-Aether theory and some variants of infrared HL gravity~\cite{librati}.
It would be interesting if one could examine what is the status of HL gravity when the extended phase space formalisim 
is applied. It could be found elsewhere.

\subsection{Non-rotating BTZ BH}
The non-rotating BTZ BH is a solution of Einstein-Maxwell gravity in three  spacetime dimension. 
The metric form is given by 
\begin{eqnarray}
ds^2 =  -\left(\frac{r^2}{\ell^2}-M \right) dt^2 
+\frac{dr^2}{\left(\frac{r^2}{\ell^2}-M \right)}+r^2d\phi^2  ~.\label{btz}
\end{eqnarray}
where $M$ is the ADM mass of the BH and $-\Lambda=\frac{1}{\ell^2}=8\pi P G_{3}$ denotes the cosmological constant.
Here we have set $c=\hbar=k=1$.
The BH OH is {located} at $r_{h}=\sqrt{8G_{3}M }\ell$~\footnote{We have already mentioned that in 
the extended phase space $\ell =\sqrt{\frac{3}{8\pi P}}$. In the subsequent expression we have to put this condition 
to obtain the results in terms of thermodynamic pressure.}. 
$G_{3}$ is 3D Newtonian constant. Interestingly, the 
thermodynamic volume for 3D static BTZ BH  computed in~\cite{dolan11}
\begin{eqnarray}
{\cal V}_{h} &=& \pi r_{h}^2=8\pi G_{3}M \ell^2
\end{eqnarray}
This is an isolated case and the thermodynamic volume is mass dependent thus it is not quantized 
{as well as it is} not universal. Where $\Lambda=-\frac{1}{\ell^2}$ is cosmological constant.

\subsection{Schwarzschild-AdS BH}
This BH is a solution of Einstein equation. The form of the metric function is given by 
\begin{eqnarray}
{\cal Z}(r) &=&  1-\frac{2M}{r}+\frac{r^2}{\ell^2}, ~\label{h3}
\end{eqnarray}
where $\Lambda=-\frac{3}{\ell^2}$ is cosmological constant. The horizon radii could be calculated from the 
following equation
\begin{eqnarray}
r^3+\ell^2 r-2M\ell^2 &=& 0  ~.\label{h4}
\end{eqnarray}
Among the three roots, only one root is real. Therefore the BH posessess only one physical horizon which is located  
at
\begin{eqnarray}
{r}_{ h} &=& \left(\frac{\ell}{3}\right)^\frac{2}{3}\left(9M+\sqrt{3}\sqrt{\ell^2+27M^2}\right)^\frac{1}{3} 
-\left(\frac{\ell^4}{3}\right)^\frac{1}{3} \frac{1}{\left(9M+\sqrt{3}\sqrt{\ell^2+27M^2}\right)^\frac{1}{3}}
~.\label{vsa}
\end{eqnarray}

The thermodynamic volume is computed to be 
\begin{eqnarray}
{\cal V}_{h} = \frac{4}{3}\pi r_{h}^3=\frac{4}{3}\pi 
\left[\left(\frac{\ell}{3}\right)^\frac{2}{3}\left(9M+\sqrt{3}\sqrt{\ell^2+27M^2}\right)^\frac{1}{3} 
-\left(\frac{\ell^4}{3}\right)^\frac{1}{3} \frac{1}{\left(9M+\sqrt{3}\sqrt{\ell^2+27M^2}\right)^\frac{1}{3}} \right]^3
~.\label{vsa1}
\end{eqnarray} 
Since it is an isolated case and the thermodynamic volume is mass dependent therefore it is not universal nor does it 
quantized.  We do not considered the other AdS spacetime because it has already been discussed in~\cite{ppmpd}.

\subsection{RN-AdS BH}
For this BH, the metric function is given by 
\begin{eqnarray}
 {\cal Z}(r) &=&  1-\frac{2M}{r}+\frac{Q^2}{r^2}+\frac{r^2}{\ell^2}, ~\label{h5}
\end{eqnarray}
The horizon radii could be found from the following equation
\begin{eqnarray}
r^4+\ell^2 r^2-2M \ell^2r+\ell^2 Q^2 &=& 0   ~.\label{h6}
\end{eqnarray}
Among the four roots, two roots are real and two roots are imaginary. Thus the OH and IH radii becomes 
$$
{r}_{h,c} = \frac{1}{2} \sqrt{\frac{1}{3} \left(\frac{x}{2} \right)^\frac{1}{3}+\left(\frac{2}{x} \right)^\frac{1}{3}
\frac{\ell^2(\ell^2+12Q^2)}{3}-\frac{2\ell^2}{3}} \pm
$$
\begin{eqnarray}
 \frac{1}{2}\sqrt{\frac{4M\ell^2}{\sqrt{\frac{1}{3} \left(\frac{x}{2} \right)^\frac{1}{3}+\left(\frac{2}{x} \right)^\frac{1}{3}
\frac{\ell^2(\ell^2+12Q^2)}{3}-\frac{2\ell^2}{3}}}-\frac{1}{3} \left(\frac{x}{2} \right)^\frac{1}{3}- 
\left(\frac{2}{x} \right)^\frac{1}{3}\frac{\ell^2(\ell^2+12Q^2)}{3}-\frac{4\ell^2}{3}}
~.\label{h7}
\end{eqnarray}
where 
$$
x=2\ell^6+108M^2\ell^4-72\ell^4Q^2+\sqrt{(2\ell^6+108M^2\ell^4-72\ell^4Q^2)^2-4\ell^6(\ell^2+12Q^2)^3}
$$
The thermodynamic volume product of RN-AdS BH for OH  and IH  is  computed to be  
\begin{eqnarray}
{\cal V}_{h} {\cal V}_{c} = \frac{\pi^2}{36} \left[\frac{2}{3} \left(\frac{x}{2} \right)^\frac{1}{3}
+\left(\frac{2}{x} \right)^\frac{1}{3}\frac{2\ell^2(\ell^2+12Q^2)}{3}+\frac{2\ell^2}{3}-
\frac{4M\ell^2}{\sqrt{\frac{1}{3} \left(\frac{x}{2} \right)^\frac{1}{3}+\left(\frac{2}{x} \right)^\frac{1}{3}
\frac{\ell^2(\ell^2+12Q^2)}{3}-\frac{2\ell^2}{3}}} \right]^3   ~.\label{h8}
\end{eqnarray} 
It is clearly evident from the above expression that the product is strictly mass-dependent. Thus the product is not 
universal. But below we would like to determine that \emph{somewhat complicated function of inner and 
outer horizon volume} is indeed mass-independent. To proceed it we would like to use the Vieta's theorem. 
Therefore from Eq.~(\ref{h6}), we get 
\begin{eqnarray}
\sum_{i=1}^{4} r_{i} &=& 0 ~.\label{h9}\\
\sum_{1\leq i<j\leq 4} r_{i}r_{j} &=& \ell^2 ~.\label{h10}\\
\sum_{1\leq i<j<k\leq 4} r_{i}r_{j} r_{k} &=&  2M\ell^2 ~.\label{h11}\\
\prod_{i=1}^{4} r_{i} &=&  \ell^2Q^2 ~.\label{h12}
\end{eqnarray}
Hence the mass-independent \emph{volume sum} and \emph{volume product} relations are
\begin{eqnarray}
\sum_{i=1}^{4} {{\cal V}_{i}}^{\frac{1}{3}} &=& 0 ~.\label{eq5}\\
\sum_{1\leq i<j\leq 4}({\cal V}_{i}{\cal V}_{j})^{\frac{1}{3}}  &=& \left(\frac{3}{32 \pi}\right)^{\frac{1}{3}}
\frac{8\pi\ell^2}{3} ~.\label{eq6}\\
\prod_{i=1}^{4} ({\cal V}_{i})^{\frac{1}{3}}&=&  
\left(\frac{\pi}{6}\right)^{\frac{1}{3}}\frac{8\pi\ell^2Q^2}{3} ~.\label{eq7}
\end{eqnarray}
The mass independent volume functional relation in terms of two horizons are 
\begin{eqnarray}
f({\cal V}_{h}, {\cal V}_{c}) &=& \ell^2 ~. \label{eq8}
\end{eqnarray}
where
\begin{eqnarray}
f({\cal V}_{h}, {\cal V}_{c}) &=&
\left(\frac{3}{32 \pi}\right)^{\frac{1}{3}} \frac{\left(\frac{8\pi\ell^2Q^2}{3}\right)}{({\cal V}_{h} {\cal V}_{c})^{\frac{1}{3}}}
-\left(\frac{3}{4\pi}\right)^{\frac{2}{3}}\left[{{\cal V}_{h}}^{\frac{2}{3}}+{{\cal V}_{c}}^{\frac{2}{3}}+({\cal V}_{h}
{\cal V}_{c})^{\frac{1}{3}}
\right] ~. \label{eq9}
\end{eqnarray}
These are explicitly mass-independent volume functional relations in the extended phase space. 

\subsection{Ho\v{r}ava Lifshitz-AdS BH}
The metric function for Ho\v{r}ava Lifshitz BH in AdS space~\cite{mun,ppGal} is given by 
\begin{eqnarray}
{\cal Z}(r) &=&  1+\left(1-\frac{2\Lambda}{3\omega} \right)\omega r^2-\omega r^2
\sqrt{1-\frac{4\Lambda}{3\omega}+\frac{4M}{\omega r^3}} ~.\label{hla}
\end{eqnarray}
The horizon radii could be calculated from the following equation
\begin{eqnarray}
4r^4+2\left(\omega\ell^2+2\right)\ell^2r^2-4M\omega\ell^4 r+\ell^4 &=& 0  ~.\label{hla1}
\end{eqnarray}
Similarly, among the four roots, two roots are real and two roots are imaginary. 
Thus the OH and IH radii becomes 
\begin{eqnarray}
{r}_{h} =\frac{\left(a+b\right)}{2},\,\, {r}_{c} =\frac{\left(a-b\right)}{2} ~.\label{hla2}
\end{eqnarray}
where 
$$
a=\sqrt{\frac{2^\frac{1}{3}
\left(\omega^2\ell^8 +4\omega\ell^6 +16\ell^4 \right)-\frac{\ell^2}{3}\left(\omega\ell^2+2\right)}
{3y^\frac{1}{3}}+ \frac{1}{12} \left(\frac{y}{2}\right)^\frac{1}{3}}
$$

$$
b= \sqrt{\frac{2M\omega\ell^4}{a}-\frac{1}{12} \left(\frac{y}{2}\right)^\frac{1}{3}- 
\frac{\left[2^\frac{1}{3}\left(\omega^2\ell^8 +4\omega\ell^6 +16\ell^4 \right)+\frac{2}{3}\ell^2
\left(\omega\ell^2+2\right)\right]}{3y^\frac{1}{3}}}
$$
and 
$$
y=1728M^2\omega^2\ell^8-576\ell^6\left(\omega\ell^2+2\right)+16\ell^6\left(\omega\ell^2+2\right)^3+
$$
\begin{eqnarray}
\sqrt{\left[1728 M^2\omega^2\ell^8-576\ell^6\left(\omega\ell^2+2\right)+16\ell^6\left(\omega\ell^2+2\right)^3\right]^2
-256\ell^{12}\left(\omega^2\ell^4+4\omega\ell^2+16 \right)^3} ~.\label{hla3}
\end{eqnarray}
The thermodynamic volume for this BH is quite different from RN-AdS spacetime and it has been calculated in~\cite{ppGal}
\begin{eqnarray}
{\cal V}_{h} &=& \frac{4}{3}\pi r_{h}^3\left[\frac{4}{\ell^2}+\frac{2}{\omega r_{h}^2} \right]  ~.\label{hla4}
\end{eqnarray} 
and 
\begin{eqnarray}
{\cal V}_{c} &=& \frac{4}{3}\pi r_{c}^3\left[\frac{4}{\ell^2}+\frac{2}{\omega r_{c}^2} \right]  ~.\label{hla5}
\end{eqnarray} 
The volume product is calculated to be 
$$
{\cal V}_{h} {\cal V}_{c} = \frac{16\pi^2}{9}\left[\frac{2^\frac{4}{3}\left(\omega^2\ell^8 
+4\omega\ell^6 +16\ell^4 \right)+\frac{\ell^2}{3}\left(\omega\ell^2+2\right)}{3y^\frac{1}{3}} 
+\frac{1}{6} \left(\frac{y}{2}\right)^\frac{1}{3}-\frac{2M\omega\ell^4}{a} \right] \times
$$
\begin{eqnarray}
\left[\frac{1}{4\ell^4}\left\{ \frac{2^\frac{4}{3}\left(\omega^2\ell^8 
+4\omega\ell^6 +16\ell^4 \right)+\frac{\ell^2}{3}\left(\omega\ell^2+2\right)}{3y^\frac{1}{3}} 
+\frac{1}{6} \left(\frac{y}{2}\right)^\frac{1}{3}-\frac{2M\omega\ell^4}{a} \right\}^2+\frac{1}{\omega^2}
+\frac{1}{\omega\ell^2} \left\{\frac{2M\omega\ell^4}{a}-\frac{\ell^2\left(\omega\ell^2+2\right)}{3y^\frac{1}{3}}\right\}
\right]   ~.\label{hla6}
\end{eqnarray} 
From the above expression we can conclude that the volume product for  Ho\v{r}ava Lifshitz BH in AdS space 
is strictly mass dependent. Thus this product is not a universal quantity. Below we will derive 
\emph{more complicated function of inner and outer horizon volume} is 
indeed mass-independent. To compute it we should apply the Vieta's theorem. Thus from Eq.~(\ref{hla1}), we find
\begin{eqnarray}
\sum_{i=1}^{4} r_{i} &=& 0 ~.\label{heq1}\\
\sum_{1\leq i<j\leq 4} r_{i}r_{j} &=& \frac{\omega \ell^4}{2}\left(1+\frac{2}{\omega\ell^2}\right)~.\label{heq2}\\
\sum_{1\leq i<j<k\leq 4} r_{i}r_{j} r_{k} &=& M \omega\ell^4 ~.\label{heq3}\\
\sum_{1\leq i<j<k<l\leq 4} r_{i}r_{j} r_{k}r_{l} &=&  \frac{\ell^4}{4} ~.\label{heq4}
\end{eqnarray}
Eliminating the mass parameter in terms of two horizons one could obtain the following 
mass-independent volume functional relation 
\begin{eqnarray}
g({\cal V}_{h}, {\cal V}_{c}) &=& 
\frac{\omega \ell^4}{2}\left(1+\frac{2}{\omega\ell^2}\right) ~,\label{heq8}
\end{eqnarray}
where
\begin{eqnarray}
g({\cal V}_{h}, {\cal V}_{c})
 &=& \frac{\left(\frac{\ell^4}{4} \right)}{r_{h}r_{c}}-\left(r_{h}^2+r_{c}^2 +r_{h}r_{c}\right)
 ~,\label{heq8.1}
\end{eqnarray}
where the parameters $r_{h}$ and $r_{c}$ could be obtain by solving the Eq.~(\ref{hla4}) and Eq.~(\ref{hla5}) 
in terms of thermodynamic volume as   
\begin{eqnarray}
r_{h} &=& \frac{1}{2} \left[\left(\frac{u_{h}}{9}\right)^\frac{1}{3}\frac{1}{\omega}-
\frac{2\ell^2}{(3u_{h})^\frac{1}{3}} \right] ~,\label{heq9}\\
r_{c} &=& \frac{1}{2} \left[\left(\frac{u_{c}}{9}\right)^\frac{1}{3}\frac{1}{\omega}-
\frac{2\ell^2}{(3u_{c})^\frac{1}{3}}\right] ~,\label{heq10}
\end{eqnarray}
and
\begin{eqnarray}
u_{h} &=& \sqrt{3}\sqrt{8\omega^3\ell^6+27\ell^4\omega^6+27\ell^4\omega^6\left(\frac{3V_{h}}{4\pi}\right)^3}
-9\ell^2\omega^3\left(\frac{3V_{h}}{4\pi}\right)~,\label{heq11}\\
u_{c} &=& \sqrt{3}\sqrt{8\omega^3\ell^6+27\ell^4\omega^6+27\ell^4\omega^6\left(\frac{3V_{c}}{4\pi}\right)^3}
-9\ell^2\omega^3\left(\frac{3V_{c}}{4\pi}\right)~.\label{heq12}
\end{eqnarray}
Now the Eq.~(\ref{heq8}) is completely mass independent volume functional relation.

\subsection{Thermodynamic Volume Products for Phantom BHs}
In this section, we would like to discuss the thermodynamic volume products for phantom AdS BH~\cite{hq}. The fact that 
phantom fields are exotic fields in BH physics. It could be generated via negative energy density. It could 
explain the acceleration of our universe. Thus one could expected that this exotic fields might have an important 
role in BH thermodynamics. We want to study here what is the key role of this phantom fields in thermodynamic 
volume functional relation? This is the main motivation behind this work. 
For phantom BH, the metric function is given by 
\begin{eqnarray}
 {\cal Z}(r) &=&  1-\frac{2M}{r}-\frac{\Lambda}{3}r^2+\eta\frac{Q^2}{r^2}, ~\label{p1}
\end{eqnarray}
where the parameter $\eta$ determines the nature of electro-magnetic~(EM) field. For $\eta=1$, one obtains the 
classical EM theory but when $\eta=-1$, one obtains the Maxwell field which is \emph{phantom}. 

Therefore for phantom BH, the horizon radii could be found from the following equation
\begin{eqnarray}
r^4+\ell^2 r^2-2M \ell^2r-\ell^2 Q^2 &=& 0   ~.\label{p2}
\end{eqnarray}
The above equation has four roots, among them the two roots are real and other two roots are imaginary. 
Thus the OH and IH radii are 
$$
{r}_{h,c} = \frac{1}{2} \sqrt{\frac{1}{3} \left(\frac{z}{2} \right)^\frac{1}{3}+\left(\frac{2}{z} \right)^\frac{1}{3}
\frac{\ell^2(\ell^2-12Q^2)}{3}-\frac{2\ell^2}{3}} \pm
$$
\begin{eqnarray}
 \frac{1}{2}\sqrt{\frac{4M\ell^2}{\sqrt{\frac{1}{3} \left(\frac{z}{2} \right)^\frac{1}{3}+\left(\frac{2}{z}\right)^\frac{1}{3}
\frac{\ell^2(\ell^2-12Q^2)}{3}-\frac{2\ell^2}{3}}}-\frac{1}{3} \left(\frac{z}{2} \right)^\frac{1}{3}- 
\left(\frac{2}{z}\right)^\frac{1}{3}\frac{\ell^2(\ell^2-12Q^2)}{3}-\frac{4\ell^2}{3}}
~.\label{p3}
\end{eqnarray}
where 
$$
z=2\ell^6+108M^2\ell^4+72\ell^4Q^2+\sqrt{(2\ell^6+108M^2\ell^4+72\ell^4Q^2)^2-4\ell^6(\ell^2-12Q^2)^3}
$$
Now we compute the thermodynamic volume product which turns out to be 
\begin{eqnarray}
{\cal V}_{h} {\cal V}_{c} = \frac{\pi^2}{36} \left[\frac{2}{3} \left(\frac{z}{2} \right)^\frac{1}{3}
+\left(\frac{2}{z}\right)^\frac{1}{3}\frac{2\ell^2(\ell^2-12Q^2)}{3}+\frac{2\ell^2}{3}-
\frac{4M\ell^2}{\sqrt{\frac{1}{3} \left(\frac{z}{2} \right)^\frac{1}{3}+\left(\frac{2}{z}\right)^\frac{1}{3}
\frac{\ell^2(\ell^2-12Q^2)}{3}-\frac{2\ell^2}{3}}}\right]^3   ~.\label{p4}
\end{eqnarray} 
The above product indicates that it is strictly mass-dependent. Therefore the product is not 
universal. Below we would like to prove that \emph{more complicated function of inner and 
outer horizon volume} is indeed mass-independent. 

To do  this we would like to use the Vieta's theorem. 
Thus from Eq.~(\ref{p2}), we find
\begin{eqnarray}
\sum_{i=1}^{4} r_{i} &=& 0 ~.\label{p5}\\
\sum_{1\leq i<j\leq 4} r_{i}r_{j} &=& \ell^2 ~.\label{p6}\\
\sum_{1\leq i<j<k\leq 4} r_{i}r_{j} r_{k} &=&  2M\ell^2 ~.\label{p7}\\
\prod_{i=1}^{4} r_{i} &=& - \ell^2Q^2 ~.\label{p8}
\end{eqnarray}
Thus the mass-independent \emph{volume sum} and \emph{volume product} relations should read
\begin{eqnarray}
\sum_{i=1}^{4} {{\cal V}_{i}}^{\frac{1}{3}} &=& 0 ~.\label{p9}\\
\sum_{1\leq i<j\leq 4}({\cal V}_{i}{\cal V}_{j})^{\frac{1}{3}}  &=& \left(\frac{3}{32 \pi}\right)^{\frac{1}{3}}
\frac{8\pi\ell^2}{3} ~.\label{p10}\\
\prod_{i=1}^{4} ({\cal V}_{j})^{\frac{1}{3}}&=&  
\left(\frac{\pi}{6}\right)^{\frac{1}{3}}\frac{8\pi\ell^2Q^2}{3} ~.\label{p11}
\end{eqnarray}
Therefore the mass independent volume functional relations in terms of two horizons are 
\begin{eqnarray}
f({\cal V}_{h}, {\cal V}_{c}) &=& -\ell^2 ~. \label{p12}
\end{eqnarray}
where
\begin{eqnarray}
f({\cal V}_{h}, {\cal V}_{c}) &=&
\left(\frac{3}{32 \pi}\right)^{\frac{1}{3}} \frac{\left(\frac{8\pi\ell^2Q^2}{3}\right)}{({\cal V}_{h} {\cal V}_{c})^{\frac{1}{3}}}
+\left(\frac{3}{4\pi}\right)^{\frac{2}{3}}\left[{{\cal V}_{h}}^{\frac{2}{3}}+{{\cal V}_{c}}^{\frac{2}{3}}+({\cal V}_{h}
{\cal V}_{c})^{\frac{1}{3}}
\right] ~. \label{p13}
\end{eqnarray}
These are explicitly mass-independent volume functional relations in the extended phase space. 

\subsection{Thermodynamic Volume Products for AdS BH in $f(R)$ Gravity}
In this section we are interested to derive the thermodynamic volume products for a static, spherically symmetric 
AdS BH in  $f(R)$ gravity. To some extent it is called modified gravity. It is a very crucial tool for explaining the  
current and  future status of the accelerating universe. Thus it is very important to investigate the thermodynamic 
volume products for this gravity.
The metric~\cite{moon11,ppmpd} function for this kind of gravity can be written as 
\begin{eqnarray}
{\cal Z}(r) &=&  1-\frac{2m}{r}+\frac{q^2}{\alpha r^2}-\frac{R_{0}}{12}r^2 ~.\label{fr}
\end{eqnarray}
where $\alpha=1+f'(R_{0})$. The parameters $m$ and $q$ are related to the ADM mass, $M$  and 
electric charge, $Q$ by the following expression
\begin{eqnarray}
m=\frac{M}{\alpha} ,\,\,\,\, q=\sqrt{\alpha}Q~. \label{fr1}
\end{eqnarray}
In this gravity, the thermodynamic pressure could be written as $P=-\frac{\Lambda}{8\pi} \alpha=\frac{3}{8\pi\ell^2}$ 
and the scalar curvature constant as $R_{0}=-\frac{12}{\ell^2}=4\Lambda$. Thus the horizon equation for $f(R)$ gravity 
becomes
\begin{eqnarray}
r^4+\ell^2 r^2-2m \ell^2r+\frac{\ell^2 q^2}{\alpha} &=& 0   ~.\label{fr2}
\end{eqnarray}
The EH radius and CH radius are
\begin{eqnarray}
{r}_{h} =\frac{\left(a+b\right)}{2},\,\, {r}_{c} =\frac{\left(a-b\right)}{2} ~.\label{fr3}
\end{eqnarray}
where 
$$
a = \sqrt{\frac{1}{3\alpha} \left(\frac{\mu}{2} \right)^\frac{1}{3}+\left(\frac{2}{\mu} \right)^\frac{1}{3}
\frac{\ell^2(\alpha\ell^2+12q^2)}{3}-\frac{2\ell^2}{3}}
$$
and 
\begin{eqnarray}
b= \sqrt{\frac{4m\ell^2}{a}-\frac{1}{3\alpha} \left(\frac{\mu}{2} \right)^\frac{1}{3}- 
\left(\frac{2}{\mu} \right)^\frac{1}{3}\frac{\ell^2(\alpha\ell^2+12q^2)}{3}-\frac{4\ell^2}{3}}
~.\label{fr4}
\end{eqnarray}
where 
$$
\mu=2\alpha^3\ell^6+108\alpha^3m^2\ell^4-72\alpha^2\ell^4q^2+
\sqrt{(2\alpha^3\ell^6+108\alpha^3m^2\ell^4-72\alpha^2\ell^4q^2)^2-4\ell^6(\alpha^2\ell^2+12\alpha q^2)^3}
$$
As is the volume products for $f(R)$  gravity  derived to be  
\begin{eqnarray}
{\cal V}_{h} {\cal V}_{c} = \frac{\pi^2}{36} \left[\frac{2}{3\alpha} \left(\frac{\mu}{2} \right)^\frac{1}{3}
+\left(\frac{2}{\mu} \right)^\frac{1}{3}\frac{2\ell^2(\alpha \ell^2+12q^2)}{3}+\frac{2\ell^2}{3}-
\frac{4m\ell^2}{a} \right]^3   ~.\label{fr5}
\end{eqnarray} 
It indicates that the volume product is not mass-independent. Now we shall give an alternative approach where we would 
see that more complicated function of volume functional relation is quite mass-independent. To derive it, we should use 
the Vieta's theorem then one could find
\begin{eqnarray}
\sum_{i=1}^{4} r_{i} &=& 0 ~.\label{fr6}\\
\sum_{1\leq i<j\leq 4} r_{i}r_{j} &=& \ell^2 ~.\label{fr7}\\
\sum_{1\leq i<j<k\leq 4} r_{i}r_{j} r_{k} &=&  2m\ell^2 ~.\label{fr8}\\
\prod_{i=1}^{4} r_{i} &=& \frac{q^2\ell^2}{\alpha } ~.\label{fr9}
\end{eqnarray}
Eliminating third and fourth roots, the mass independent volume functional relation derived as 
\begin{eqnarray}
f({\cal V}_{h}, {\cal V}_{c}) &=& \ell^2 ~. \label{fr10}
\end{eqnarray}
where
\begin{eqnarray}
f({\cal V}_{h}, {\cal V}_{c}) &=&
\left(\frac{3}{32 \pi}\right)^{\frac{1}{3}} 
\frac{\left(\frac{8\pi\ell^2q^2}{3}\right)}{\alpha ({\cal V}_{h} {\cal V}_{c})^{\frac{1}{3}}}
-\left(\frac{3}{4\pi}\right)^{\frac{2}{3}}\left[{{\cal V}_{h}}^{\frac{2}{3}}+{{\cal V}_{c}}^{\frac{2}{3}}
+({\cal V}_{h}{\cal V}_{c})^{\frac{1}{3}}\right] ~. \label{fr11}
\end{eqnarray}
This equation is explicitly mass-independent.

\subsection{Thermodynamic Volume Products for Regular BH}
In this section we compute the thermodynamic volume products for a regular BH derived by Ay\'{o}n-Beato and 
Garc\'{i}a~(ABG)~\cite{abg,ppgrg}. It is a spherically symmetric solution of Einstein's general relativity and it 
is a curvature singularity free solution. The metric function form of ABG BH is given by
\begin{eqnarray}
{\cal Z}(r) &=& 1-\frac{2mr^2}{(r^2+q^2)^{\frac{3}{2}}}+\frac{q^2r^2}{(r^2+q^2)^2} ~.\label{g1}
\end{eqnarray}
where $m$ is the mass of the BH and $q$ is the monopole charge. The horizon radii could be found from the 
following equation 
\begin{eqnarray}
r^{8}+(6q^2-4m^2)r^{6}+(11q^4-4m^2q^2)r^{4} +6q^6r^2+q^8 = 0 ~.\label{g2}
\end{eqnarray}
This is a polynomial equation of order $8^{th}$. This could be reduced to fourth order polynomial 
equation by putting $r^2=z$ then one obtains~\cite{ppgrg}
\begin{eqnarray}
z^{4}+(6q^2-4m^2)z^{3}+(11q^4-4m^2q^2)z^{2} +6q^6z+q^8 &=& 0 ~.\label{g3}
\end{eqnarray}
The EH and CH are located at
\begin{eqnarray}
r_{h} &=& \sqrt{\frac{(2m^2-3q^2)}{2}+\frac{a}{2}+\frac{b}{2}}\\
r_{c} &=& \sqrt{\frac{(2m^2-3q^2)}{2}-\frac{a}{2}-\frac{b}{2}}
\end{eqnarray}
and the other horizons~\footnote{We have considered only here EH and CH. The other horizons are discarded.} are 
located at 
\begin{eqnarray}
r_{hc} &=& \sqrt{\frac{(2m^2-3q^2)}{2}+\frac{a}{2}-\frac{b}{2}}\\
r_{ch} &=& \sqrt{\frac{(2m^2-3q^2)}{2}-\frac{a}{2}+\frac{b}{2}}
\end{eqnarray}
where 
\begin{eqnarray}
a = \sqrt{4m^2q^2-11q^4+(2m^2-3q^2)^2+\frac{(11q^4-4m^2q^2)}{3}+\left(\frac{2}{\delta}\right)^\frac{1}{3}
\frac{(16m^4q^4-16m^2q^6+25q^8)}{3}+\frac{1}{3}\left(\frac{\delta}{2}\right)^\frac{1}{3}}
\end{eqnarray}
and 
\begin{eqnarray}
b &=& \sqrt{c}
\end{eqnarray}
where
$$
c= 4m^2q^2-11q^4+2(2m^2-3q^2)^2+\frac{(4m^2q^2-11q^4)}{3}-\left(\frac{2}{\delta}\right)^\frac{1}{3}
\frac{(16m^4q^2-16m^2q^6+25q^8)}{3}-
$$
$$
\frac{1}{3}\left(\frac{\delta}{2}\right)^\frac{1}{3}+\frac{\{48q^6-8(2m^2-3q^2)^3+8(2m^2-3q^2)(11q^4-4m^2q^2)\}}{4a}
$$
and 
$$
\delta=624 m^4q^8-128m^6q^6-240m^2q^{10}+250q^{12}
$$
\begin{eqnarray}
+\sqrt{324864m^8q^{16}-110592m^{10}q^{14}-193536m^6q^{18}+172800m^4q^{20}}~.\label{g4.1}
\end{eqnarray}

The volume product of ${\cal H}^{\pm}$ is evaluated to be
\begin{eqnarray}
{\cal V}_{h} {\cal V}_{c} =\frac{\pi^2}{36} \left[(2m^2-3q^2)^2-(a+b)^2\right] ~.\label{g4}
\end{eqnarray} 
As usual, the volume product is not mass-independent. Now we would see below 
\emph{somewhat more complicated  function of inner and outer horizon volume} 
is indeed mass-independent. To derive it we have to apply Vieta's theorem in 
Eq.~(\ref{g3}) thus one obtains
\begin{eqnarray}
\sum_{i=1}^{4} z_{i} &=& 4m^2-6q^2 ~.\label{g5}\\
\sum_{1\leq i<j\leq 4} z_{i}z_{j} &=& 11q^4- 4m^2q^2  ~.\label{g6}\\
\sum_{1\leq i<j<k\leq 4} z_{i}z_{j} z_{k} &=&  -6q^2 ~.\label{g7}\\
\prod_{i=1}^{4} z_{i} &=& q^8 ~.\label{g8}
\end{eqnarray}
Eliminating the mass parameter one obtains the mass independent equation in terms of two horizons 
$$
z_{1}z_{2}(z_{1}+z_{2})+6q^2z_{1}z_{2}-q^8\frac{(z_{1}+z_{2})}{z_{1}z_{2}}-
$$
\begin{eqnarray}
\frac{1}{z_{1}+z_{2}+q^2}\left[(z_{1}+z_{2})^2+6q^2(z_{1}+z_{2})-z_{1}z_{2}-\frac{q^8}{z_{1}z_{2}}+11q^4\right] 
&=& 6q^6  ~.\label{g8.1}
\end{eqnarray}
It should be noted that the symbols $(h,c)$ and $(1,2)$ both have same meaning. Now in terms of volume of 
${\cal H}^{\pm}$ the mass-independent volume functional relation becomes 
\begin{eqnarray}
f({\cal V}_{h}, {\cal V}_{c}) &=& 6q^6 ~. \label{g9}
\end{eqnarray}
where 
$$
f({\cal V}_{h}, {\cal V}_{c})= \left(\frac{3}{4\pi}\right)^2 \left({\cal V}_{h} {\cal V}_{c}\right)^{\frac{2}{3}}
\left\{{\cal V}_{h}^{\frac{2}{3}}+{\cal V}_{c}^{\frac{2}{3}}\right\}+6q^2 \left(\frac{3}{4\pi}\right)^{\frac{4}{3}}
\left({\cal V}_{h} {\cal V}_{c}\right)^{\frac{2}{3}}
-q^8\left(\frac{4\pi}{3}\right)^{\frac{2}{3}} \frac{\left({\cal V}_{h}^{\frac{2}{3}}+{\cal V}_{c}^{\frac{2}{3}}\right)}
{\left({\cal V}_{h} {\cal V}_{c}\right)^{\frac{2}{3}}}
$$
$$
-\left(\frac{3}{4\pi}\right)^{\frac{2}{3}}\frac{\left({\cal V}_{h} {\cal V}_{c}\right)^{\frac{2}{3}}}
{\left\{{\cal V}_{h}^{\frac{2}{3}}+{\cal V}_{c}^{\frac{2}{3}}+\left(\frac{4\pi}{3}\right)^{\frac{2}{3}}q^2\right\}}\times
$$
\begin{eqnarray}
\left[\left(\frac{3}{4\pi}\right)^{\frac{4}{3}}\left({\cal V}_{h}^{\frac{2}{3}}+{\cal V}_{c}^{\frac{2}{3}}\right)^2+
6q^2\left(\frac{3}{4\pi}\right)^{\frac{2}{3}}\left({\cal V}_{h}^{\frac{2}{3}}+{\cal V}_{c}^{\frac{2}{3}}\right)
-\left(\frac{3}{4\pi}\right)^{\frac{4}{3}}\left({\cal V}_{h} {\cal V}_{c}\right)^{\frac{2}{3}} 
-\left(\frac{4\pi}{3}\right)^{\frac{2}{3}}\frac{q^8}{\left({\cal V}_{h} {\cal V}_{c}\right)^{\frac{2}{3}}}+11q^4 \right]
\end{eqnarray}
Now we are moving to axisymmetric spacetime. To see what happens there?

\section{Axisymmetric Spacetime}
In this section we have considered only the various  axisymmetric BHs. It is easy to compute volume products for 
spherically symmetric cases bacause of $ {\cal V}_{h} \propto {\cal A}_{h} r_{h}$ for OH and 
$ {\cal V}_{c} \propto {\cal A}_{c} r_{c}$ for IH. While for axisymmetric spacetime this proportionality  is quite 
different because here the spin parameter is present. Now see what happens in this case by starting with Kerr BH.

\subsection{Kerr BH}
The Kerr BH is a solution of Einstein equation. The OH radius and IH radius reads for this BH should read
\begin{eqnarray} 
r_{h}=M + \sqrt{M^2-a^2} \\
r_{c}= M - \sqrt{M^2-a^2}  ~.\label{kr1}
\end{eqnarray}
where $a=\frac{J}{M}$. $J$ is angular momentum of the BH. When $M^2>a^2$, it descibes a BH while $M^2<a^2$ it descibes 
a naked singularity. 
The thermodynamic volume for OH~\cite{nata} and IH~\cite{ppmpd} becomes 
\begin{eqnarray}
{\cal V}_{h} &=& \frac{{\cal A}_{h} r_{h}}{3} \left[1+\frac{a^2}{2r_{h}^2} \right]  \\
{\cal V}_{c} &=& \frac{{\cal A}_{c} r_{c}}{3} \left[1+\frac{a^2}{2r_{c}^2} \right] ~.\label{kr2}
\end{eqnarray} 
The thermodynamic volume product of Kerr BH for OH  and IH  is calculated to be  
\begin{eqnarray}
{\cal V}_{h} {\cal V}_{c} &=& \frac{128}{9} \pi^2 J^2 M^2 \left(1+\frac{a^2}{8M^2} \right)   ~.\label{kr3}
\end{eqnarray} 

The volume sum for OH  and IH  is  
\begin{eqnarray}
{\cal V}_{h}+ {\cal V}_{c} &=& \frac{32}{3}\pi M^3 \left(1- \frac{a^2}{4M^2} \right)   ~.\label{kr4}
\end{eqnarray} 
Similarly, the volume minus for OH  and IH is 
\begin{eqnarray}
{\cal V}_{h}- {\cal V}_{c} &=& \frac{32}{3}\pi M^2 \sqrt{M^2-a^2}    ~.\label{kr5}
\end{eqnarray} 
and the volume division is 
\begin{eqnarray}
\frac{{\cal V}_{h}}{ {\cal V}_{c}} &=& \left(\frac{4M^2-a^2 + 4M\sqrt{M^2-a^2}}{4M^2-a^2 - 4M\sqrt{M^2-a^2}}\right)  
~.\label{kr6}
\end{eqnarray} 
It indicates that the volume product, volume sum, volume minus and volume division  for Kerr BH is mass dependent. 
Therefore  the product, the sum, the minus and the division all are \emph{not} universal.

\subsection{Kerr-AdS BH}
The horizon function for Kerr-AdS BH~\cite{carter} is given by 
\begin{eqnarray}
 \Delta_{r} &=& \left(r^2+a^2 \right)\left(1+\frac{r^2}{\ell^2}\right)-2Mr =0 ~.\label{kad}
\end{eqnarray}
which gives the quartic order of horizon equation
\begin{eqnarray}
\frac{r^4}{\ell^2} +\left(1+\frac{a^2}{\ell^2}\right)r^2-2mr+a^2 &=& 0 ~.\label{kad1}
\end{eqnarray}
The quantities $m$ and $a$ are related to the parameters mass $M$ and angular momentum $J$ as follows
\begin{eqnarray}
 m &=& M \Xi^2, \,\, a=\frac{J}{m} \Xi^2
\end{eqnarray}
where $\Xi=1-\frac{a^2}{\ell^2}$. To obtain the roots of Eq.~(\ref{kad1}) we apply the Vieta's theorem, we find
\begin{eqnarray}
\sum_{i=1}^{4} r_{i} &=& 0 ~.\label{eq1}\\
\sum_{1\leq i<j\leq 4} r_{i}r_{j} &=& \ell^2 \left(1+\frac{a^2}{\ell^2}\right) ~.\label{eq2}\\
\sum_{1\leq i<j<k\leq 4} r_{i}r_{j} r_{k} &=& 2m\ell^2 ~.\label{eq3}\\
\prod_{i=1}^{4} r_{i} &=&  a^2\ell^2 ~.\label{eq4}
\end{eqnarray}
There is at least two real zeros of the Eq.~(\ref{kad1}) which is OH radius and IH radius. 
After some algebraic computation we have 
\begin{eqnarray}
r_{h}+r_{c} &=& \frac{2m\ell^2}{a^2+\ell^2+r_{h}^2+r_{c}^2}  , \label{kd2} \\
r_{h}r_{c} &=& \frac{a^2\ell^2-(r_{h}r_{c})^2}{a^2+\ell^2+r_{h}^2+r_{c}^2} ~.\label{kad2}
\end{eqnarray}
The area of the BH for OH is 
\begin{eqnarray}
{\cal A}_{h} &=& \frac{4\pi \left(r_{h}^2+a^2\right)}{\Xi}   ~.\label{kad3}
\end{eqnarray}
and for IH is 
\begin{eqnarray}
{\cal A}_{c} &=& \frac{4\pi \left(r_{c}^2+a^2\right)}{\Xi}   ~.\label{kad4}
\end{eqnarray}
The thermodynamic volume for OH~\cite{dolan11,cvetic11}  becomes 
\begin{eqnarray}
{\cal V}_{h} =  \frac{2\pi \left[(r_{h}^2+a^2)(2r_{h}^2\ell^2+a^2\ell^2-r_{h}^2 a^2)\right]}{3r_{h} \ell^2 \Xi^2}  
~.\label{kad5}
\end{eqnarray} 
and we derive the thermodynamic volume for IH  becomes
\begin{eqnarray}
{\cal V}_{c} = \frac{2\pi \left[(r_{c}^2+a^2)(2r_{c}^2\ell^2+a^2\ell^2-r_{c}^2 a^2)\right]}{3r_{c} \ell^2 \Xi^2}  
~.\label{kad6}
\end{eqnarray} 
The thermodynamic volume product for Kerr-AdS BH is calculated in Eq.~(\ref{kad7}).
\begin{eqnarray}
{\cal V}_{h} {\cal V}_{c} = \frac{4\pi^2\{r_{h}^2r_{c}^2+a^2 \left(r_{h}^2+r_{c}^2 \right)+a^4\}}{9\Xi^4 r_{h}r_{c}} \times
\left[3r_{h}^2r_{c}^2+2a^2 \left(r_{h}^2+r_{c}^2 \right)+\Xi^2 r_{h}^2r_{c}^2 \right]~.\label{kad7}
\end{eqnarray}
Using Eq.~(\ref{kd2}), Eq.~(\ref{kad2}), Eq.~(\ref{kad3}) and Eq.~(\ref{kad4}), we observe that there is no way to 
eliminate the mass parameter from Eq. (\ref{kad7}) therefore the volume product for Kerr-AdS BH is not mass-independent 
thus it is not universal and not quantized . 

\subsection{Kerr-Newman BH}
It is an axisymmetric solution of Einstein-Maxwell equations.  The OH radius and IH radius for this 
BH becomes
\begin{eqnarray} 
r_{h}=M + \sqrt{M^2-a^2-Q^2} \\
r_{c}= M - \sqrt{M^2-a^2-Q^2}  ~.\label{kn}
\end{eqnarray}
The thermodynamic volume for OH~\cite{dolan11} is 
\begin{eqnarray}
{\cal V}_{h} &=& \frac{2\pi \left[(r_{h}^2+a^2)(2r_{h}^2+a^2)+a^2Q^2\right]}{3r_{h}}  ~.\label{kn1}
\end{eqnarray} 
and we derive that the thermodynamic volume for IH is
\begin{eqnarray}
{\cal V}_{c} &=& \frac{2\pi \left[(r_{c}^2+a^2)(2r_{c}^2+a^2)+a^2Q^2\right]}{3r_{c}}  ~.\label{kn2}
\end{eqnarray} 
The thermodynamic volume product for KN BH is computed to be 
$$
{\cal V}_{h} {\cal V}_{c} = \frac{16\pi^2}{9} \times
$$
\begin{eqnarray}
\frac{\left[J^2(8J^2+a^4-a^2Q^2-2Q^4+8M^2Q^2)+Q^4(a^2+Q^2)^2 \right]}{a^2+Q^2}  
~.\label{kn3}
\end{eqnarray}
It also indicates that the thermodynamic volume for KN BH is mass dependent. Thus the volume product 
is not universal for any axisymmetric spacetime. In the appropriate limit i.e. when  $a=J=0$, one obtains 
the thermodynamic volume product for Reissner Nordstr\"{o}m BH and  when $Q=0$, one obtains the volume product 
for Kerr BH.

\subsection{Kerr-Newman-AdS BH}
The horizon function for Kerr-Newman-AdS BH~\cite{calda} reads
\begin{eqnarray}
 \Delta_{r} &=& \left(r^2+a^2 \right)\left(1+\frac{r^2}{\ell^2}\right)-2Mr+q^2 =0 ~.\label{knad}
\end{eqnarray}
which has the quartic order of horizon equation
\begin{eqnarray}
\frac{r^4}{\ell^2} +\left(1+\frac{a^2}{\ell^2}\right)r^2-2mr+a^2+q^2 &=& 0 ~.\label{knad1}
\end{eqnarray}
The quantity $q$ is related to the charge parameter $Q$ as
\begin{eqnarray}
 q &=& Q \Xi 
\end{eqnarray}
To determine the roots of Eq.~(\ref{knad1}) again we apply the Vieta's rule then one obtains
\begin{eqnarray}
\sum_{i=1}^{4} r_{i} &=& 0 ~.\label{kq1}\\
\sum_{1\leq i<j\leq 4} r_{i}r_{j} &=& \ell^2 \left(1+\frac{a^2}{\ell^2}\right) ~.\label{kq2}\\
\sum_{1\leq i<j<k\leq 4} r_{i}r_{j} r_{k} &=& 2m\ell^2 ~.\label{kq3}\\
\prod_{i=1}^{4} r_{i} &=&  (a^2+q^2)\ell^2 ~.\label{kq4}
\end{eqnarray}
Similarly, there is at least two real zeros of the Eq.~(\ref{knad1}) which is OH radius and IH radius. 
After some algebraic derivation one get
\begin{eqnarray}
r_{h}+r_{c} &=& \frac{2m\ell^2}{a^2+\ell^2+r_{h}^2+r_{c}^2}  , \label{knd2} \\
r_{h}r_{c} &=& \frac{(a^2+q^2)\ell^2-(r_{h}r_{c})^2}{a^2+\ell^2+r_{h}^2+r_{c}^2} ~.\label{knad2}
\end{eqnarray}
The area of this BH for OH is 
\begin{eqnarray}
{\cal A}_{h} &=& \frac{4\pi \left(r_{h}^2+a^2\right)}{\Xi}   ~.\label{knad3}
\end{eqnarray}
and for IH is 
\begin{eqnarray}
{\cal A}_{c} &=& \frac{4\pi \left(r_{c}^2+a^2\right)}{\Xi}   ~.\label{knad4}
\end{eqnarray}
The thermodynamic volume for OH~\cite{dolan11,cvetic11}  becomes 
\begin{eqnarray}
{\cal V}_{h} = \frac{2\pi \left[(r_{h}^2+a^2)(2r_{h}^2\ell^2+a^2\ell^2-r_{h}^2 a^2)
+\ell^2q^2a^2\right] }{3r_{h} \ell^2 \Xi^2} 
~.\label{knad5}\nonumber 
\end{eqnarray} 
and we derive the thermodynamic volume for IH  becomes
\begin{eqnarray}
{\cal V}_{c} = \frac{2\pi \left[(r_{c}^2+a^2)(2r_{c}^2\ell^2+a^2\ell^2-r_{c}^2 a^2)
+\ell^2q^2a^2\right] }{3r_{c} \ell^2 \Xi^2} 
~.\label{knad6}\nonumber 
\end{eqnarray} 
The thermodynamic volume product for Kerr-Newman-AdS BH is computed in Eq.~(\ref{knad7}).
$$
{\cal V}_{h} {\cal V}_{c} = \frac{4\pi^2}{9\Xi^4 r_{h}r_{c}} \times
$$
$$
\left[ \{r_{h}^2r_{c}^2+a^2 \left(r_{h}^2+r_{c}^2 \right)+a^4\}^2+
\{r_{h}^2r_{c}^2+a^2 \left(r_{h}^2+r_{c}^2 \right)+a^4\} \{2r_{h}^2r_{c}^2+a^2 \left(r_{h}^2+r_{c}^2 \right)\}\right]
$$
$$
+\frac{4\pi^2 \left[a^2q^2 \{r_{h}^4+r_{c}^4+2a^2 \left(r_{h}^2+r_{c}^2 \right)+2a^4\} 
+\Xi^2 r_{h}^2r_{c}^2 \{r_{h}^2r_{c}^2+a^2 \left(r_{h}^2+r_{c}^2 \right)+a^4\}\right]}{9\Xi^4 r_{h}r_{c}} 
$$
\begin{eqnarray}
+\frac{4\pi^2 \left[a^2q^2\Xi \{r_{h}^4+r_{c}^4+a^2 \left(r_{h}^2+r_{c}^2 \right)\}+a^4q^4\right]}{9\Xi^4 r_{h}r_{c}}  
~.\label{knad7}
\end{eqnarray}
Again using Eq.~(\ref{knd2}), Eq.~(\ref{knad2}), Eq.~(\ref{knad3}) and Eq.~(\ref{knad4}), we speculate that there has been 
no chance to eliminate the mass parameter from Eq. (\ref{knad7}) thus the volume product for Kerr-Newman-AdS BH is not 
mass-independent therefore it is not universal and not quantized.

\subsection{Spinning BTZ BH}
The metric for rotating BTZ BH~\cite{btz92} in $2+1$ dimension is given by
$$
ds^2 =  -\left(\frac{r^2}{\ell^2}+\frac{J^2}{4r^2}-M \right) dt^2 
+\frac{dr^2}{\left(\frac{r^2}{\ell^2}+\frac{J^2}{4r^2}-M \right)}
$$
\begin{eqnarray}
+r^2\left(-\frac{J}{2r^2} dt+d\phi\right)^2  ~.\label{btzm}
\end{eqnarray}
where $M$ and $J$ represents the ADM mass, and the angular momentum of the BH. 
$-\Lambda=\frac{1}{\ell^2}=8\pi P G_{3}$ denotes the cosmological constant.
Here we have set $8G_{3}=1=c=\hbar=k$. When $J=0$, one obtains the static BTZ BH.

The BH OH radius and IH radius are~\cite{btz92,ppjetpl}
\begin{eqnarray} 
r_{h} &=& \sqrt{\frac{M \ell^2}{2}\left(1+ \sqrt{1-\frac{J^2}{M^2 \ell^2}} \right)} .~\label{s1}\\
r_{c} &=& \sqrt{\frac{M \ell^2}{2}\left(1- \sqrt{1-\frac{J^2}{M^2 \ell^2}} \right)} .~\label{s2}
\end{eqnarray}
The thermodynamic volume for 3D spinning BTZ BH  for OH and IH:
\begin{eqnarray}
{\cal V}_{h} &=& \left(\frac{\partial M}{\partial P}\right)_{J}= \pi r_{h}^2\\
{\cal V}_{c} &=& \left(\frac{\partial M}{\partial P}\right)_{J}= \pi r_{c}^2
\end{eqnarray}
The thermodynamic volume product is computed to be 
\begin{eqnarray}
{\cal V}_{h} {\cal V}_{c} &=&  \frac{\pi^2 J^2 \ell^2}{4}    ~.\label{s3}
\end{eqnarray} 
Interestingly, the thermodynamic volume product for rotating BTZ BH is \emph{mass-independent i.e. universal}
and it is also quantized. This is the only example for \emph{rotating} cases, the volume product is universal.
This is an interesting result of this work.

\section{Discussion}
In this work, we have demonstrated that the thermodynamic products in particular thermodynamic volume 
products of spherically symmetric spacetime and axisymmetric spacetime by incorporating the extended 
phase-space formalism. In this formalism, the cosmological constant should be considered as a thermodynamic
pressure and its conjugate parameter as thermodynamic volume. In addition to that the mass parameter should be 
treated as enthalpy of the system rather than internal energy. Then in this phase space the first law of BH 
thermodynamics should be satisfied for both the OH and IH.

We explicitly computed the thermodynamic volume products both for OH and IH of several classes 
of spherically symmetric and axisymmetric BHs including the AdS spacetime. In this cases, the simple 
volume product of ${\cal H}^{\pm}$ is not mass independent. Rather slightly more complicated volume 
functional relations are indeed mass-independent. We have proved that for simple  Reissner Nordstr\"{o}m BH
of Einstein gravity and Kehagias-Sfetsos BH of Ho\v{r}ava Lifshitz gravity, the thermodynamic volume product 
of ${\cal H}^{\pm}$ is indeed \emph{universal}. Such products are \emph{mass independent} 
for spherically symmetric cases  because of ${\cal V}_{h} \propto {\cal A}_{h} r_{h}$ for 
OH and ${\cal V}_{c} \propto {\cal A}_{c} r_{c}$ for IH. 

Axisymmetric spacetime does not satisfied this proportionality due to presence of the spin parameter thus 
such spacetime shows \emph{no} mass-independent features except the rotating BTZ BH, the only axisymmetric 
spacetime in $3D$ showed that \emph{universal} features thus it has been quantized in this sense. We also computed 
thermodynamic volume sum but they are always mass dependent so they are not universal as well as they are not quantized.

Like area~(or entropy) products, the simple thermodynamic volume product of ${\cal H}^{\pm}$ is not mass-independent rather 
more complicated function of volume functional relation is indeed mass independent. This is often true for spherically 
symmetric BHs including AdS spacetime. This senerio for axisymmetric spacetime ~(except 3D BTZ BH ) is quite different. 
In this cases, the area functional relation becomes mass-independent whereas the volume functional relation is not 
mass-independent. For volume products, this is the main differences between spherically symmetric spacetime and 
axisymmetric spacetime. To sum up, the volume functional relation that we have studied in this work 
in spherically symmetric cases~(but not for axisymmetric cases) further provides some  universal properties 
of the BH which gives some insight of microscopic origin of BH entropy both outer and inner.

\end{document}